\documentclass{PoS}

\usepackage[english]{babel}
\usepackage{amsmath}
\usepackage[comma,square,numbers,sort]{natbib}
\usepackage{array}

\newcommand{\eVq}  {\text{eV}^2}

\newcommand{\Dmq}  {\Delta m^2}
\newcommand{\Dms}  {\Delta m^2_{21}}
\newcommand{\Dma}  {\Delta m^2_{31}}

\title{Three-flavour neutrino oscillation update and comments on
possible hints for a non-zero $\theta_{13}$}

\ShortTitle{Three-flavour neutrino oscillation update}

\author{Michele Maltoni\\
        Departamento de F\'isica Te\'orica \& Instituto de F\'isica
	Te\'orica UAM/CSIC, Facultad de Ciencias C-XI, Universidad
	Aut\'onoma de Madrid, Cantoblanco, E-28049 Madrid, Spain\\
	E-mail: \email{michele.maltoni AT uam.es}}

\author{\speaker{Thomas Schwetz}\\
        Max-Planck-Institute for Nuclear Physics,
        PO Box 103980, 69029 Heidelberg, Germany\\
        E-mail: \email{schwetz AT mpi-hd.mpg.de}}

      \abstract{We provide a short summary of three-flavour neutrino
        oscillation parameters as determined from a global fit to 2008
        data, and we comment on possible hints in favour of a non-zero
        value of the mixing angle $\theta_{13}$ found in
        arXiv:0806.2649.  We do confirm a hint from solar + KamLAND
        data at about 1.5$\sigma$, which can be understood from the
        recent SNO CC/NC measurment. However, we show that a claimed
        hint from atmospheric data is much less robust. It depends on
        details of event rate calculations and treatment of
        theoretical uncertainties.  We could identify two data points
        showing an ``excess'' (at the 1$\sigma$ level) in the SK-I
        multi-GeV $e$-like data, which seem to be the origin of the
        slight preference for $\theta_{13} > 0$. We point out that
        once SK-I and SK-II data are combined this ``excess''
        disappears, and irrespective of the details of the analysis,
        no ``hint'' from atmospheric data is obtained for the SK-I and
        SK-II combined data set. As a result the global fit of all
        data leads to a best fit value of $\theta_{13}$ consistent
        with zero within less than 1$\sigma$.}

\FullConference{Identification of dark matter 2008\\
		 August 18-22, 2008\\
		 Stockholm, Sweden}

\begin{document}

\section{Introduction}
\label{sec:introduction}

Thanks to the synergy amongst a variety of experiments involving solar
and atmospheric neutrinos, as well as man-made neutrinos at nuclear
power plants and accelerators neutrino physics has undergone a
revolution over the last decade or so by establishing the phenomenon
of neutrino oscillations. The parameters relevant for three-flavour
neutrino oscillations are three mixing angles,
$\theta_{12},\theta_{13},\theta_{23}$, and one CP-phase in the unitary
lepton mixing matrix, and two mass-squared differences $\Dms \equiv
m^2_2-m^2_1$ and $\Dma \equiv m^2_3- m^2_1$. In this note we summarize
the determination of these parameters from present world neutrino
oscillation data (Sec.~\ref{sec:summary}), and we give a critical
discussion of recent claims for possible hints in favour of a non-zero
value of the mixing angle $\theta_{13}$ (Sec.~\ref{sec:th13}).

\section{Summary of three-flavour  oscillation parameters}
\label{sec:summary}

Table~\ref{tab:summary} summarizes the results of two recent global
fits to world neutrino data from Refs.~\cite{Schwetz:2008er,
  GonzalezGarcia:2007ib}.  For another recent analysis
see~\cite{Fogli:2008ig}.

\begin{table}[ht]\centering
\begin{tabular}{|@{\quad}l@{\quad}|@{\quad}c@{\quad}@{\quad}c@{\quad}|@{\quad}c@{\quad}@{\quad}c@{\quad}|}
        \hline
        & \multicolumn{2}{c|@{\quad}}{Ref.~\cite{Schwetz:2008er}}
        & \multicolumn{2}{c|}{Ref.~\cite{GonzalezGarcia:2007ib}
	  (MINOS updated)}
        \\
        parameter
	& best fit$\pm 1\sigma$ & 3$\sigma$ interval
	& best fit$\pm 1\sigma$ & 3$\sigma$ interval
        \\
        \hline
        $\Delta m^2_{21}\: [10^{-5}\eVq]$
        & $7.65^{+0.23}_{-0.20}$ & 7.05--8.34
	& $7.67^{+0.22}_{-0.21}$ & 7.07--8.34
	\\[1mm]
        $\Delta m^2_{31}\: [10^{-3}\eVq]$
        & $\pm 2.40^{+0.12}_{-0.11}$ & \hspace{-9pt}$\pm$(2.07--2.75)
	&
	\begin{tabular}{c}
	    $-2.39 \pm 0.12$ \\
	    $+2.49 \pm 0.12$
	\end{tabular}
	& \hspace{-11pt}
	\begin{tabular}{c}
	    $-$(2.02--2.79) \\
	    $+$(2.13--2.88)
	\end{tabular}
	\\[4mm]
        $\sin^2\theta_{12}$
        & $0.304^{+0.022}_{-0.016}$ & 0.25--0.37
        & $0.321^{+0.023}_{-0.022}$ & 0.26--0.40
	\\[2mm]
        $\sin^2\theta_{23}$
        & $0.50^{+0.07}_{-0.06}$ & 0.36--0.67
	& $0.47^{+0.07}_{-0.06}$ & 0.33--0.64 
	\\[2mm]
        $\sin^2\theta_{13}$
        & $0.01^{+0.016}_{-0.011}$ & $\leq$ 0.056
        & $0.003\pm 0.015$ & $\leq$ 0.049
	\\
        \hline
\end{tabular}
\caption{\label{tab:summary}%
  Determination of three--flavour neutrino oscillation parameters from
  2008 global data~\cite{Schwetz:2008er, GonzalezGarcia:2007ib}.}
\end{table}

\begin{figure}
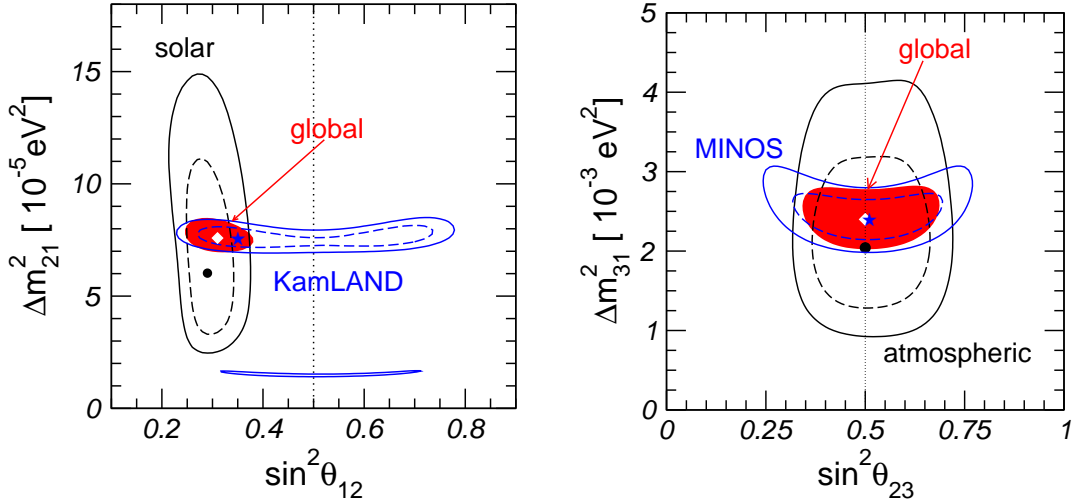

 \centering
  \includegraphics[height=0.44\textwidth]{F-sol-kl-08-3nu.eps}\qquad
  \includegraphics[height=0.44\textwidth]{F-atm-min-08-3nu.eps}
  \caption{Determination of the leading ``solar'' and ``atmospheric''
    oscillation parameters~\cite{Schwetz:2008er}. We show allowed
    regions at 90\% and 99.73\%~CL (2~dof) for solar and KamLAND
    (left), and atmospheric and MINOS (right), as well as the
    99.73\%~CL regions for the respective combined analyses.}
  \label{fig:dominant}
\end{figure}

The latest data release from the KamLAND reactor
experiment~\cite{:2008ee} has increased the exposure almost fourfold
over previous results~\cite{Araki:2004mb} to a total exposure of
2881~ton$\cdot$yr and also systematic uncertainties have been
improved.
The Sudbury Neutrino Observatory (SNO) has released the data of its
last phase, where the neutrons produced in the neutrino neutral
current (NC) interaction with deuterium are detected mainly by an
array of $^3$He NC detectors (NCD)~\cite{Aharmim:2008kc}. The main
impact of the new SNO data is due to the lower value for the observed
CC/NC ratio, $(\phi_\mathrm{CC}/\phi_\mathrm{NC})^\mathrm{NCD} =
0.301\pm0.033$~\cite{Aharmim:2008kc}, compared to the previous value
$(\phi_\mathrm{CC}/\phi_\mathrm{NC})^\mathrm{salt} =
0.34\pm0.038$~\cite{Aharmim:2005gt}.  Since for $^8$B neutrinos
$\phi_\mathrm{CC}/\phi_\mathrm{NC} \approx P_{ee} \approx
\sin^2\theta_{12}$, adding the new data point on this ratio with the
lower value leads to a stronger upper bound on $\sin^2\theta_{12}$.
This explains the slightly lower best fit point for
$\sin^2\theta_{12}$ found in~\cite{Schwetz:2008er} compared
to~\cite{GonzalezGarcia:2007ib}, since the latter does not (yet)
include the SNO-NCD result.
Data from SNO are combined with the global data on solar
neutrinos~\cite{solar-ex} and with the recent results from
Borexino~\cite{Arpesella:2008mt}, reporting a survival probability of
the 0.862 MeV $^7$Be neutrinos of $P_{ee}^\mathrm{^7Be,obs} = 0.56 \pm
0.1$.
Fig.~\ref{fig:dominant}(left) illustrates how the determination of the
leading solar oscillation parameters $\theta_{12}$ and $\Dmq_{21}$
emerges from the complementarity of solar and reactor neutrinos.
Spectral information from KamLAND data leads to an accurate
determination of $\Delta m^2_{21}$ with the remarkable precision of
8\% at $3\sigma$, defined as $(x^\mathrm{upper} -
x^\mathrm{lower})/(x^\mathrm{upper} + x^\mathrm{lower})$. KamLAND
data start also to contribute to the lower bound on
$\sin^2\theta_{12}$, whereas the upper bound is dominated by solar
data, most importantly by the CC/NC solar neutrino rates measured by
SNO.

The MINOS experiment has reported new results on $\nu_\mu$
disappearance with a baseline of 735~km based on a two-year exposure
from the Fermilab NuMI beam corresponding to a total of $3.36 \times
10^{20}$ protons on target~\cite{Adamson:2008zt}. The data confirm the
energy dependent disappearance of $\nu_\mu$, showing significantly
less events than expected in the case of no oscillations in the energy
range $\lesssim 6$~GeV, whereas the high energy part of the spectrum
is consistent with the no oscillation expectation. We combine the
long-baseline accelerator data from MINOS as well as from
K2K~\cite{Aliu:2004sq} with atmospheric neutrino measurements from
Super-Kamiokande~\cite{Ashie:2005ik}, using the results of
Ref.~\cite{Maltoni:2004ei}. In this analysis sub-leading effects of
$\Dms$ in atmospheric data are neglected, but effects of $\theta_{13}$
are included.
Fig.~\ref{fig:dominant}(right) illustrates how the determination of
the leading atmospheric oscillation parameters $\theta_{23}$ and
$|\Dmq_{31}|$ emerges from the complementarity of atmospheric and
accelerator neutrino data.  The determination of $|\Dma|$ is dominated
by spectral data from the MINOS experiment, where the sign of
$\Dmq_{31}$ (i.e., the neutrino mass hierarchy) is undetermined by
present data. The measurement of the mixing angle $\theta_{23}$ is
still largely dominated by atmospheric neutrino data from
Super-Kamiokande with a best fit point close to maximal mixing. Small
deviations from maximal mixing due to sub-leading three-flavour
effects (not included in the analysis
of~\cite{Schwetz:2008er,Maltoni:2004ei}) have been found in
Refs.~\cite{Fogli:2005cq,GonzalezGarcia:2007ib}, {\it c.f.}\
Tab.~\ref{tab:summary}.  See, however, also Ref.~\cite{suzuki} for a
preliminary analysis of Super-Kamiokande. A comparison of these subtle
effects can be found in Ref.~\cite{snow}. At present deviations from
maximality are not statistically significant.

\begin{figure}
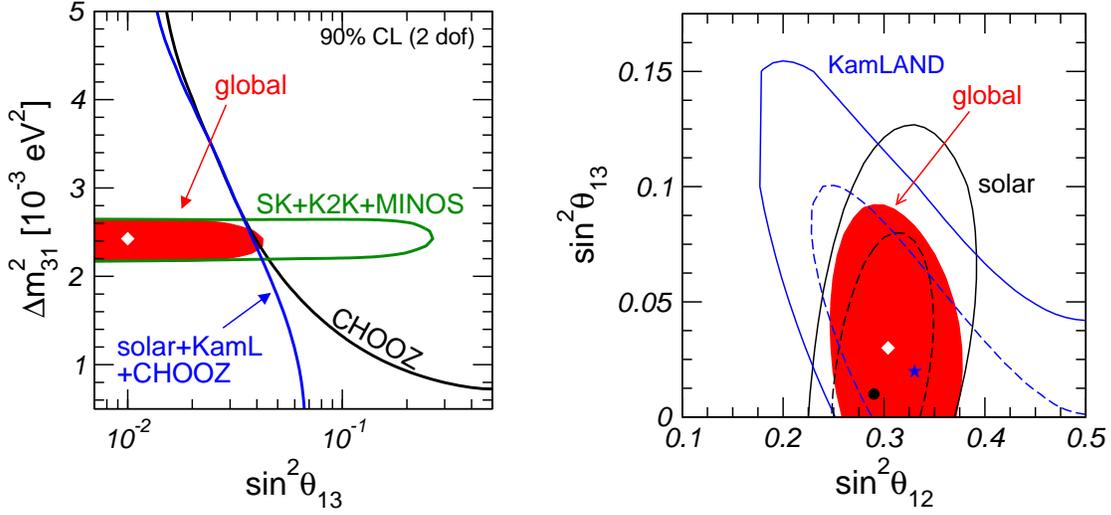

  \centering
  \includegraphics[height=0.45\textwidth]{th13-2008-lin.eps}\qquad
  \includegraphics[height=0.44\textwidth]{s12-s13-tension.eps}
  \caption{Left: Constraints on $\sin^2\theta_{13}$ from the interplay
    of different parts of the global data. Right: Allowed regions in
    the $(\theta_{12}-\theta_{13})$ plane at 90\% and 99.73\%~CL
    (2~dof) for solar and KamLAND, as well as the 99.73\%~CL region
    for the combined analysis. $\Dms$ is fixed at its best fit
    point. The dot, star, and diamond indicate the best fit points of
    solar, KamLAND, and combined data, respectively.}
  \label{fig:th13}
\end{figure}

The third mixing angle $\theta_{13}$ would characterize the magnitude
of CP violation in neutrino oscillations and is crucial also for the
determination of the neutrino mass ordering. It is the main objective
of upcoming reactor and accelerator experiments to measure this
parameter.
Fig.~\ref{fig:th13}(left) summarizes the information on $\theta_{13}$
from present data, which emerges from an interplay of different data
sets. An important contribution to the bound comes, of course, from
the CHOOZ reactor experiment~\cite{Apollonio:2002gd} combined with the
determination of $|\Dmq_{31}|$ from atmospheric and long-baseline
experiments. The results on $\theta_{13}$~\cite{Schwetz:2008er}
reported in Tab.~\ref{tab:summary} imply an upper bound of
$\sin^2\theta_{13} < 0.035 (0.056)$ at 90\% CL (3$\sigma$) and the
best fit point is consistent with zero within $0.9\sigma$, while
Ref.~\cite{GonzalezGarcia:2007ib} finds the slightly stronger bound
$\sin^2\theta_{13} < 0.049$ at 3$\sigma$ and a $\Delta\chi^2 = 0.04$
for $\theta_{13} = 0$. These differences can be understood by noting
that SNO-NCD data is not included in the results
from~\cite{GonzalezGarcia:2007ib}, see below for an explanation.

\section{Comments on possible hints for a non-zero 
value of $\theta_{13}$}
\label{sec:th13}

A recent global analysis of neutrino data~\cite{Fogli:2008jx} obtains
a hint for a non-zero value of $\theta_{13}$ at 1.6$\sigma$, which
emerges from a 1.2$\sigma$ hint from solar+KamLAND data combined with
a 0.9$\sigma$ hint from atmospheric+long-baseline+CHOOZ data. The hint
from solar+KamLAND data can be understood by the slight downward shift
of the SNO CC/NC ratio due to the SNO-NCD data mentioned in
Sec.~\ref{sec:summary}. From the combination of solar
and KamLAND data we find a best fit value of $\sin^2\theta_{13} =
0.03$ with $\Delta\chi^2 = 2.2$ for $\theta_{13} = 0$ which
corresponds to a $1.5\sigma$ effect (86\%~CL). 
We illustrate the interplay of solar and KamLAND data in
Fig.~\ref{fig:th13}(right). The relevant survival probabilities are
given by
\begin{equation}\label{eq:Pee}
P_{ee} \approx
\left\{\begin{array}{l@{\qquad}l}
\cos^4\theta_{13} \left(1- \sin^22\theta_{12}\right \langle \sin^2\phi\rangle) & 
\mbox{solar, low energies / KamLAND} \\
\cos^4\theta_{13} \, \sin^2\theta_{12} & 
\mbox{solar, high energies} 
\end{array}\right. \,,
\end{equation}
where $\phi = \Dms L / 4E$ and $\langle \sin^2\phi\rangle \approx 1/2$
for solar neutrinos.  Eq.~(\ref{eq:Pee}) implies an anti-correlation
of $\sin^2\theta_{13}$ and $\sin^2\theta_{12}$ for KamLAND and low
energy solar neutrinos. In contrast, for the high energy part of the
spectrum, which undergoes the adiabatic MSW conversion inside the sun
and which is subject to the SNO CC/NC measurement, a positive
correlation of $\sin^2\theta_{13}$ and $\sin^2\theta_{12}$ emerges. As
visible from Fig.~\ref{fig:th13}(right) and as discussed already
in~\cite{Maltoni:2004ei,Goswami:2004cn}, this complementarity leads to
a non-trivial constraint on $\theta_{13}$ and it allows to understand
the hint for a non-zero value of $\theta_{13}$, which helps to
reconcile the slightly different best fit points for $\theta_{12}$ (as
well as for $\Dms$, see Fig.~\ref{fig:dominant}) for solar and KamLAND
separately~\cite{Balantekin:2008zm,Goswami:2004cn,Fogli:2008jx,Maltoni:2003da}.
This trend was visible already in pre-SNO-NCD data, though with a
significance of only $0.8\sigma$, see v6 of~\cite{Maltoni:2004ei}.

\begin{figure}
  \centering
  \includegraphics[width=0.95\textwidth]{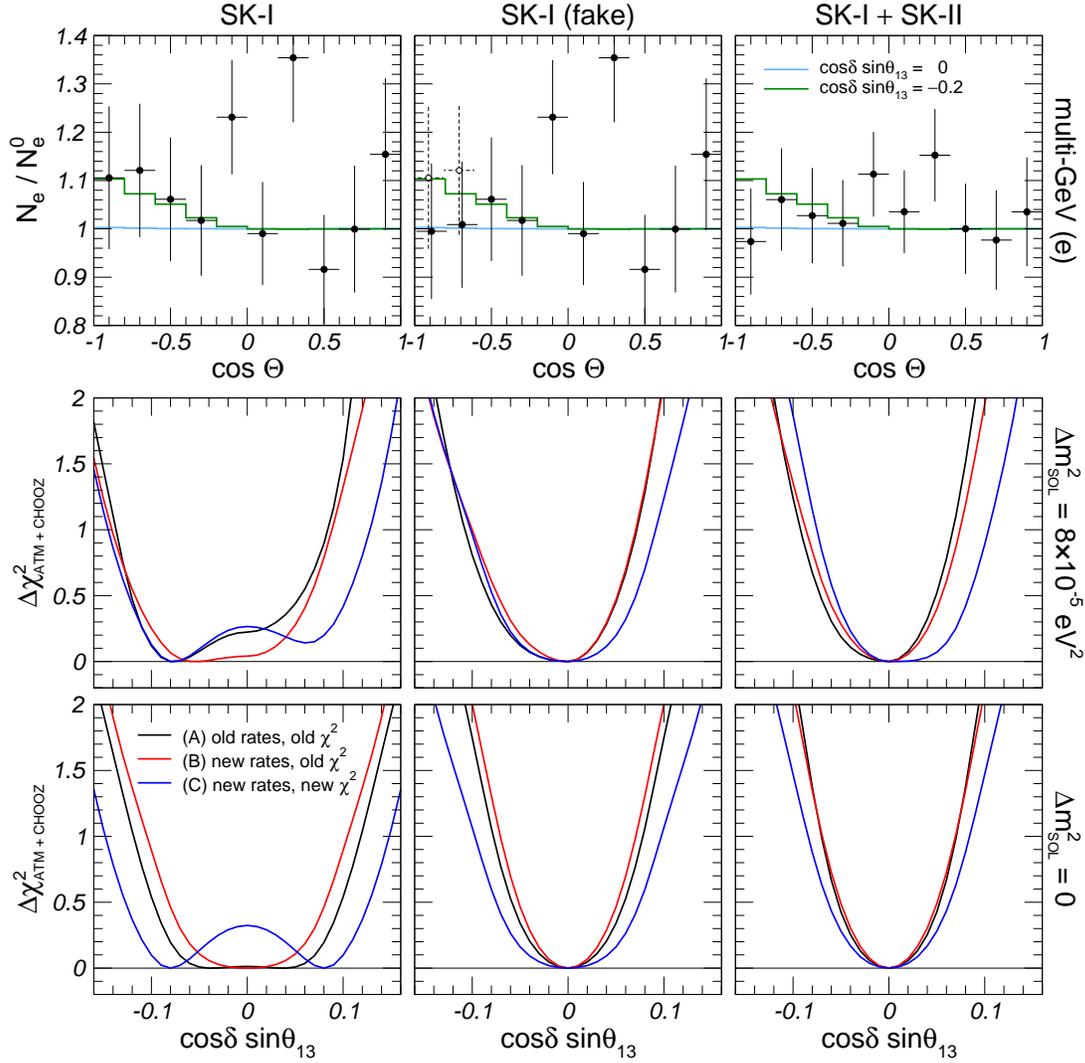}
  \caption{\label{fig:th13-atm}%
    Zenith distribution for multi-GeV $e$-like events (upper panels), 
    and $\Delta\chi^2$ dependence on $\cos\delta \sin\theta_{13}$ for
    $\Dmq_{21} = 8\times 10^{-5}~\eVq$ (middle panels) and for
    $\Dmq_{21} = 0$ (lower panels). The fits include the CHOOZ data as
    well as the full atmospheric data samples for SK-I (left panels),
    SK-I with modified multi-GeV $e$-like data (central panels) and
    SK-I + SK-II (right panels). The black, red and blue lines
    correspond to different analyses. See text for details.}
\end{figure}

The hint arising from atmospheric data is, in our opinion, much more
controversial. This preference for a non-zero $\theta_{13}$ value was
first noted in Ref.~\cite{Fogli:2005cq}, and is illustrated in Fig.~24
of that paper (see also~\cite{Escamilla:2008vq}). Our attempts to
reproduce that figure with our own numerical codes are shown in the
left panels of Fig.~\ref{fig:th13-atm}. We have performed three
different kinds of calculations:
\begin{enumerate}
  \item[(A)] {\it old rates, old $\chi^2$} (black line). This case
    follows closely the analysis presented in
    Ref.~\cite{Maltoni:2004ei}, except for the fact that here we have
    also included sub-leading $\Dmq_{21}$ effects. Note that the data
    samples and the $\chi^2$ definition used here are the same as in
    Ref.~\cite{Fogli:2005cq}; in particular, SK-I data are divided
    into 55 energy and zenith bins.
    
  \item[(B)] {\it new rates, old $\chi^2$} (red line). This case
    differs from the previous one for the details of the event rate
    calculations, which are now performed as described in the appendix
    of Ref.~\cite{GonzalezGarcia:2007ib}; on the other hand, the
    $\chi^2$ definition and the number of data points are unchanged.

  \item[(C)] {\it new rates, new $\chi^2$} (blue line). This case
    fully coincides with the three-neutrino analysis presented in
    Ref.~\cite{GonzalezGarcia:2007ib}. The treatment of the
    theoretical uncertainties in the $\chi^2$ definition differs
    considerably from the previous two cases, and the number of data
    points is now increased to 90.
\end{enumerate}
For definiteness we consider only normal hierarchy, and we have
explicitly verified that the inclusion of K2K and/or MINOS does not
affect the results, compare also~\cite{Fogli:2005cq} (without MINOS)
and~\cite{Fogli:2008jx} (with MINOS). Our analysis shows that:
\begin{itemize}
  \item the results of case (A) are qualitatively very similar to
    those shown in Fig.~24 of Ref.~\cite{Fogli:2005cq}. In particular,
    the data favor $\cos\delta = -1$ and disfavor $\cos\delta = +1$,
    indicating that a relevant contribution to the effect comes from
    the interference term $\Delta_3$ (see Eq.~44 of
    Ref.~\cite{Fogli:2005cq}). Indeed, from the lower-left panel we
    see that the hint for non-zero $\theta_{13}$ disappears for
    $\Dmq_{21} = 0$, in agreement with \cite{Fogli:2005cq};
    
  \item on the other hand, the \emph{strength} of the preference for
    non-zero $\theta_{13}$ in our analysis (A) is only $0.5\sigma$,
    much weaker than the $0.9\sigma$ found in
    Ref.~\cite{Fogli:2005cq}. Since the $\chi^2$ definition and the
    experimental data are the same for both analyses, we conclude that
    the discrepancy must arise from differences in the rate
    calculations. Indeed, a comparison between our cases (A) and (B)
    in Fig.~\ref{fig:th13-atm} shows that the statistical relevance of
    the signal strongly depends on the details of the Monte-Carlo: for
    example, for case (B) the hint of non-zero $\theta_{13}$ is only
    at the $0.2\sigma$ level, {\it i.e.}\ practically inexistent;
    
  \item details of the data binning and the theoretical uncertainties
    are also very important, as can be seen by comparing cases (B) and
    (C), which are based on identical rate calculations. In case (C)
    the preference for non-zero $\theta_{13}$ is again as strong as in
    case (A), but now the hint persists also for $\Dmq_{21} = 0$.
\end{itemize}

It is quite difficult to trace the preference for non-zero
$\theta_{13}$ to the effect of a single data sample. Clearly, the
asymmetry between the two values $\cos\delta = \pm 1$ shows that the
interference term $\Delta_3$, mostly relevant for sub-GeV data, plays
an important role. On the other hand, other samples are affected by
$\theta_{13}$, and as we have seen the details of the statistical
analysis determine whether the hint appears also when the interference
term is suppressed ({\it e.g.}, for $\Dmq_{21} = 0$), or not.
As an example, in the upper panels of Fig.~\ref{fig:th13-atm} we show
the contribution of multi-GeV $e$-like data to the effect. It is clear
from this plot that the first two bins of this sample are better
fitted with a non-zero value of $\theta_{13}$. To assess the relevance
of this contribution, in the middle and right panels of
Fig.~\ref{fig:th13-atm} we repeat the calculations described above for
a different choice of experimental data:
\begin{itemize}
  \item in the middle panels, we artificially reduce the first two
    bins of SK-I multi-GeV $e$-like events by 10\%. As can be seen,
    with this simple change the hint in favor of non-zero
    $\theta_{13}$ completely disappears, \emph{irrespective} of the
    details of the rate calculations and of the statistical analysis,
    and also of the inclusion of sub-leading $\Dmq_{21}$ effects. We
    conclude therefore that the excess in these two bins somewhat
    ``triggers'' the hint for non-zero $\theta_{13}$;
    
  \item in the right panels, we repeat our fits for the combination of
    SK-I and SK-II data. As can be seen from the upper panel, the
    ``excess'' in the first bins of multi-GeV $e$-like data no longer
    appears once SK-II data are also included. The lower panels show
    that the results are very similar to the ``fake'' SK-I data
    displayed in the middle panels, and in particular no preference
    for non-zero $\theta_{13}$ is present.
\end{itemize}
In the light of these results, we conclude that (1) the statistical
relevance of the claimed hint for non-zero $\theta_{13}$ from
atmospheric data depends strongly on the details of the rate
calculations and of the $\chi^2$ analysis, and (2) the hint is the
result of a statistical fluctuation in SK-I data which disappears
after the inclusion of SK-II data. 
Since in the atmospheric neutrino analysis used in
\cite{Schwetz:2008er,Maltoni:2004ei} no hint for $\theta_{13} > 0$ is
present, the indication from solar data becomes weakened in the global
analysis, leading to the $\theta_{13}$ result quoted in
Tab.~\ref{tab:summary}, consistent with zero at 0.9$\sigma$. We
believe that the present ``indication'' for a non-zero $\theta_{13}$ should
not be taken more serious than a $\sim 1\sigma$ effect (which of
course implies that it will be true with a probability of $\sim 68\%$)
and has to wait for being confirmed or refuted by upcoming experiments.

\paragraph{Acknowledgments.} 
TS is supported by the European Community-Research Infrastructure
Activity under the FP6 ``Structuring the European Research Area''
program (CARE, contract number RII3-CT-2003-506395).
MM is supported by MICINN through the Ram\'on y Cajal program and
through the national project FPA2006-01105, and by the Comunidad
Aut\'onoma de Madrid through the HEPHACOS project P-ESP-00346.  MM
acknowledges the hospitality of The Abdus Salam ICTP where this work
was completed.

\end{document}